Fabiano Minni [*]

A.I.F. History of Physics Group – Ferrara, Italy

# Limiting velocity and generalized Lorentz transformations

## Summary

After a short *Historical bibliographical note*, in the *Starting points* attention will be focused on some postulates common to classical mechanics and special relativity.

Starting from these premises, in the sections *The deduction of the form of possible transformations* and *The Ignatowsky constant* it will be shown that the choice between the Galilean scheme and one of the generalized Lorentz type is in fact the only possible one.

In a generalized Lorentz scheme, the interactions propagate at a finite velocity $V_L$ and the form of the transformations of the space-time coordinates of the events is analogous to those of Lorentz; the only difference is that the limiting speed $V_L$ plays the role assumed by the invariance of c, the speed of light in a vacuum, in Lorentz transformations.

The proposed derivation articulates in detail the aspects related to calculation, in order to encourage an in-depth study also aimed at upper secondary school students. The line of development of the arguments does not depend directly on electromagnetism, in other words we are dealing with a regulating principle for kinematics and for the set of laws of physics. The assumption $V_L=c$ can be assumed from the experimental datum of the invariance of c, i.e. from the validity of Maxwell's equations.

Galileo transformations are obtained *if and only if* the time interval Δt between two events is an invariant for inertial frames of reference, if this interval is not invariant then the transformations are of the generalized Lorentz type.

In this framework, the experimental confirmations of the non-invariance of Δt constitute an indirect confirmation of the generalized Lorentz transformations and therefore of the existence of a limiting velocity for the interactions.

During the course of the discussion, a demonstration of the *"Reciprocity Lemma"* will also be presented, different and simpler than other approaches proposed in the literature [6,7,8].

## Historical bibliographic notes

The first deduction of Lorentz transformations, an alternative to that of A. Einstein in 1905 [1], was made in 1910 by W. A. von Ignatowsky [2], who did not use the principle of constancy of the speed of light in a vacuum, but based his analysis only on the principle of relativity, on the isotropy and homogeneity of space, on the homogeneity of time. The following is assumed to be self-evident *principle of reciprocity*: if v is the speed with which the inertial system S sees the inertial system S' move then S' sees S move at velocity -v. Ignatowsky shows the existence of a universal constant that has the dimensions of the inverse square of a velocity but the question of its value and its sign is not addressed theoretically:

*".. It follows that n (which we can denote as a universal space-time constant) is the reciprocal of the square of a velocity, therefore a positive quantity. We see that we have obtained transformation equations similar to Lorentz's, except that n is used instead of $1/c^2$. However, the sign is still*

---

[*] E-mail : fmdragan282@gmail.com

*undetermined... To determine the sign and numerical value of n we must keep in mind the experiment."* [2].

Among the in-depth studies in the period before the Second World War, those of P. Frank-H. Rothe of 1911 [3] are worth mentioning, which take up the work of Ignatowsky, of L.A.Pars of 1921 [4], of V.Lalan of 1937 [5]; all assume the principle of reciprocity without demonstrating it.

In the Pars article it is emphasized that the hypothesis of homogeneity and isotropy of space and homogeneity of time is actually limited to situations in which the effects, characteristic of general relativity, of gravitational fields can be neglected; In general, transformations are non-linear. Pars then justifies the assumption of the principle of reciprocity with the need that the systems of measurement of S and S' must be coordinated and by convention he posits that the relative velocity measured by S and S' is the same, apart from the sign.

V. Lalan formalizes that transformations must be linear and deepens their group structure. The argument of the principle of causality against the hypothesis of a negative value of Ignatowsky's constant is then introduced: if the constant assumed a negative value then the principle of causality would be violated (for a formulation of the principle of causality, see § 1.4 .

Since the 50s of the last century, the theory has been further refined and formalized. First of all, we cite the contribution of C. Cattaneo in 1958 [6] in which the terms used and the conditions on the functions involved in the transformation formulas are specified. Then, with a hypothetical deductive method based on the principle of kinematic relativity, the homogeneity of space-time and the isotropy of space, we arrive at generalized Lorentz transformations. As already for V. Lalan, the negative value of the Ignatowsky constant is considered unacceptable because it violates the principle of causality. Cattaneo also manages to show how the principle of reciprocity can be demonstrated and not simply assumed intuitively, the principle of reciprocity becomes the lemma of reciprocity. This result was then obtained by others, for example by V. Benzi and V. Gorini in 1969 [7] and by J.M.Levy-Leblond in 1976 [8].

Almost all the works do not mention Cattaneo, whose contribution, with rare exceptions such as G. Giuliani and I. Bonizzoni [9] and G. Giuliani [10], is mostly forgotten.

In the work of Benzi and Gorini, see also the bibliography contained therein, against the negative value of Ignatowsky's constant, in addition to the appeal to the principle of causality, arguments related to the transformation of velocities are raised. By composing two positive velocities it is possible to obtain a negative one under certain conditions, something appears paradoxical but admissible in principle, and what is more serious there are cases in which there are singularities [see App.2].

Levy-Leblond's article represents an in-depth critical reflection and is also accessible as an in-depth study to high school students in the last year of the course. It should be emphasized that the assumptions of homogeneity of space-time and isotropy of space can be questioned in approaches that go beyond the theory of special relativity (think, for example, of the theory of general relativity). It is then shown that a negative value of the universal constant introduced by Ignatowsky is incompatible with the principle of causality.

An interesting line of development to arrive at generalized Lorentz transformations can be found in a work by B. Coleman of 2003 [11] (see also the bibliography contained therein).

In particular, Coleman analyses the possible values of the Ignatowsky constant and points out that negative values are to be excluded due to the occurrence of singularities in the composition of velocities [see App.2]. An interesting article from 2015 by A. Pellissetto and A. Testa [35], accessible as an in-depth study for high school students, elegantly shows how Galileo's transformations and those that generalize Lorentz transformations are the only ones compatible with the principle of relativity and with the request that the transformations constitute a group.

As for more formalized approaches that use mathematical tools that are not always within the reach of high school students but which constitute a good framework for synthesis and in-depth analysis, reference can be made to the articles by S. Cacciatori, V. Gorini, A. Kamenshchik in 2008 [12] and by Y. Friedman and T. Scarr in 2019 [36] and the bibliography contained therein.

# 1. The starting points

## § 1.1 Space is Euclidean, homogeneous and isotropic, time is homogeneous

It is assumed, similarly to what happens in classical mechanics, that space is Euclidean, the relations between the positions of bodies can be described in terms of Euclidean geometry.

The postulate of the homogeneity and isotropy of space means that all points and all directions are equivalent from the point of view of the laws of physics; The postulate of the homogeneity of time states that the laws of physics remain unchanged over time.

In essence, the result of an experience, if the environmental conditions remain the same, does not change depending on the time and place in which it is carried out and the chosen place is invariant with the direction.

These assumptions may seem "natural", in reality the principle of the uniformity of nature has gradually been affirmed only since the Copernican revolution and has been consolidated with the development of Newtonian mechanics, in particular celestial mechanics. Before then, the current opinion, following Aristotle, was that of a division between sublunar physics and that of the celestial spheres, a Universe that was anything but homogeneous and isotropic.

With special relativity, the geometry of space remains Euclidean, space retains the characteristics of homogeneity and isotropy and time that of homogeneity, but there is no reciprocal independence of space and time as in classical mechanics. Space and time are connected in a four-dimensional manifold [ref.16], Minkowski's space-time, in which "*... space in itself and time in itself must fall into darkness and only a kind of union of the two must preserve its individuality...*" [25].

A characteristic point of special relativity is the insistence on the need for all inertial reference frames, SRIs, to use units of measurement for lengths defined in an analogous way and to be able to construct a space-time grid in which clocks synchronized in the same way are associated with each point in space.

From the point of view of general relativity, however, it is necessary to speak of space-time-matter, the contribution of gravity to the structure of space-time becomes inescapable; Minkoswski's space-time is only a local approximation. In this regard, we quote the words of S.Bergia [26].

"*... around each point, or rather each event, we can choose a frame of reference with respect to which the laws of physics take the form predicted by special relativity... the geometry of the surroundings is Minkowskian...*" and again "*... Special relativity is a theory of the continuous space-time in the absence of gravitation... General relativity will be a theory of the space-time continuum in the presence of gravitation, as well as a theory of gravitation, which will incorporate special relativity in the sense of making it valid only locally...*"

On a global scale, the geometry of space-time is deformed, with respect to the absence of matter, by gravity and the geodesics of space-time no longer coincide with the lines of Euclidean geometry; the hypotheses of homogeneity and isotropy are however valid locally.

This last observation, combined with a greater simplicity in dealing with space-time intervals and velocities, will lead us to choose to treat the transformation of coordinates in terms of the transformation of their differentials.

## §1.2 Central role of inertial reference frames, the principle of relativity

In general, every reference system S defines "event" as what happens at some point P at a certain instant of time, and associates to each particular event E a real numerical quatern (x,y,z,t), hence the identification E≡(x,y,z,t).

Both in classical mechanics and in special relativity we start from the concepts of velocity, uniform rectilinear motion, inertial reference frame.

It is interesting to observe how the concept of speed already relates space and time, which is not as immediate as it appears; For a long time, the relationship between non-homogeneous quantities was thought of as something to be avoided.

For example, for Galileo [13] the comparison between two speeds necessarily passes through two distinct relationships between spaces and times (the original text is in Latin, the following translation appears in [14]):

"*Attention. Is the motion for the vertical AD faster or not faster than that for the inclined AB ? It seems so; in fact, equal spaces are travelled more quickly on AD than on AB. And yet it also seems not; in fact [...] the time for AB is to the time for AC as AB is to AC; therefore the moments of velocities for AB and for AC are the same; it is in fact one and the same speed that, at different times, travels different spaces, but having the same proportion of times.*"

Translated into modern terms, called $V_2$ and $V_1$ two speeds, we have that:

$$\frac{V_2}{V_1} = \frac{S_2}{S_1} \cdot \frac{t_1}{t_2}$$

It is only at the end of the 1600 and in the first decades of the 1700, with the development of differential calculus and analytical mechanics, that mathematical physics was born and the concept of velocity, in particular instantaneous velocity, was consolidated (see, for example, the contributions of P. Varignon, to whom we owe the first definition of instantaneous velocity, J. Bernoulli, J.B.D'Alembert, L. Euler [15]) . More generally, the relationships between non-homogeneous quantities are treated from a formal point of view and new concepts are introduced, which allows the construction of representations of the physical world within increasingly complex mathematical structures.

Another common element between classical mechanics and special relativity is the postulation, independently of the homogeneity and isotropy of space and the homogeneity of time, that there are reference frames in which the motion of a body subject to overall zero forces is a uniform rectilinear motion; these systems constitute the set $\Omega$ of inertial reference frames.

This is undoubtedly a problematic postulate that we do not intend to go into here, for a discussion of it see for example [16, 17, 18] and related bibliography.

However, we underline two direct consequences:

- the postulation of the existence of the class of inertial frames of reference is equivalent to formulating the
  "Principle of inertia" both in classical mechanics and in special relativity;
- a uniform rectilinear motion in one inertial system is still a uniform rectilinear motion in another inertial system.

Reasoning on the whole $\Omega$ we can at this point formulate the "principle of relativity", whose *empirical basis* is that experiments conducted under similar conditions in different *approximately inertial reference frames* lead to *similar experimental results*.

It is assumed that the relationships between the physical quantities, i.e. the physical laws, that appear in any experiment are invariant in all SRI ; the results of an experiment do not depend on the particular SRI in which the experiment is carried out and therefore are not able to discriminate between the various possible SRI.

In essence, all inertial systems are equivalent and it then becomes central to determine in the passage from one system to another a class of transformations of the coordinates of the events that can guarantee this equivalence, i.e. the invariance of the laws of physics.

The equivalence of inertial frames does not mean, however, that they are indistinguishable.

Each $S \in \Omega$ is in principle able to represent another S' of $\Omega$ which moves with uniform rectilinear motion at a speed $\vec{v}$ with respect to it, it is sufficient for this purpose that it measures from its point of view the speed of a point at rest with respect to S' .

If for some physical reason we attribute to an SU the role of privileged reference (think for example of the cosmic microwave background) then, however, the point of view of the other SRI must be equivalent for the description of physical phenomena. The line of reasoning presented here remains valid even if we expand the class of reference frames, for example in general relativity [16].

The history of ideas on the relativity of motions up to Galileo is a long one [18] that starts at least from the first century B.C. with Lucretius [19], passes through medieval physics with the School of Paris, in particular J. Buridan and N. d'Oresme with his *Livre du ciel e du mond* of 1377, up to the work of Giordano Bruno *The Supper of Ashes,* published in London in 1583 and very similar to the famous passage of the second day of the *Dialogue on the highest systems*, published by Galileo in 1632 [20].

To move from empirical observations on the relativity of motions to a well-structured principle, it is necessary to wait for the formulations of H. Poincaré in 1904 and A. Einstein in 1905.

*"The laws of physical phenomena must be the same for both a fixed observer and an observer transported in uniform translational motion; so that we have not and will not be able to have any means of discerning whether or not we are carried in such motion*." H. Poincaré [21] .

We remain in the field of classical mechanics, the "fixed" observer is the one who is integral with the ether that maintains the role of privileged reference. In order to justify the invariance of the speed of light in a vacuum, it is necessary to find transformations that leave Maxwell's equations unchanged in the transition from the aether system to that of any SRI. The Galilean transformation equations were not good, others had to be sought; H.A. Lorentz formulated the solution, which was later perfected by Poincaré and taken up by Einstein.

Despite the correctness of the analysis of the concept of simultaneity defined by the emission and reception of light signals, the reference to a "true" time linked to a hypothetical absolute reference makes the time "seen" by the various SRIs declassify the rank of "local time".

"*For all coordinate systems for which the equations of mechanics apply, the same electrodynamic and optical laws will also apply. We will elevate this conjecture (the content of which will be called, in what follows, "principle of relativity") to the rank of a postulate."* A. Einstein [22].

There is no longer a fixed reference, all SRI are perfectly equivalent and the evaluations of spatial and temporal intervals are equally legitimate. The introduction of the postulate of the invariance of the speed of light in a vacuum becomes a regulating principle for the construction of transformations and represents an indissoluble bond between space and time.

The transformations obtained by Einstein coincide with those formalized by Lorentz and Poincaré but physics changes radically.

### § 1.3 The set of coordinate transformations is a group

For simplicity, let's take as a spatial reference a trio of orthogonal Cartesian axes and consider the set of boosts along the axis X. Let $S_0$ and $S_1$ be two SRI , $S_0$ sees $S_1$ moving at speed $v_1$ ; we denote by $T_{01}(v_1)$ the corresponding boost. Let $S_2$ now be a third SRI moving at speed $v_2$ with respect to $S_1$ and we denote with $T_{12}(v_2)$ the corresponding boost.

Let's assume that there is a velocity $v_3$ which corresponds to a boost $T_{02}(v_3)$ that makes you go from $S_0$ to $S_2$, in other words the composition " $\cdot$ " of $T_{01}$ and $T_{12}$ generates a boost still belonging to T:

$$T_{12}\left(v_2\right) \cdot T_{01}\left(v_1\right) = T_{02}\left(v_3\right) \in T$$

The algebraic structure $(T, \cdot)$ is therefore closed. The boost with $v=0$ corresponds to the identity transformation, $T(0) = I$.

The transformations are invertible, i.e. for every $T_{ab}$ there exists $T_{ba}$ such that $T_{ab} \cdot T_{ba} = I$ . With the request for the validity of the associative property $(T_1 \cdot T_2) \cdot T_3 = T_1 \cdot (T_2 \cdot T_3)$ for every $T_1$ , $T_2$ , $T_3$ belonging to T the algebraic structure $(T, \cdot)$ becomes a group.

We emphasize the importance from a physical point of view of the closure condition, the lack of which implies the impossibility of passing from the representation of a certain event in an SRI to the representation of the same event in some SRI.

Without the closure condition it is difficult to think that we can speak of equivalence between all inertial systems and in particular that we can formulate the principle of relativity.

For an introduction to group theory in special relativity, see for example [16] and for a more general framework [23, 24].

### §1.4 The principle of causality

Without going into the meanders of the discussion on the concept of cause, we will say that a necessary but not sufficient condition for an event $E_1$ *can be* the cause of an event $E_2$ is that $E_1$ precedes $E_2$ in time ; the *causal connectivity* presupposes an increasing temporal ordering from $E_1$ to $E_2$ and this ordering must be invariant.

Principle of causality*: there is at least one non-empty class of events that preserves the temporal order according to the representation of each inertial frame of reference.*

Let's take some examples by choosing two arbitrary SRI, S and S'.

From the point of view of S, $E_1$ is the passage of a particle at point $P_1$ at time $t_1$ and $E_2$ is the arrival of the same particle at point $P_2$ at time $t_2$ with $t_2 > t_1$. According to S' the same particle passes through the point $P'_1$ at time $t'_1$ , event $E'_1$ , and $E'_2$ is the arrival of the particle at point $P'_2$ at time $t'_2$ , where the quantities with quotes are the corresponding ones of the respective quantities without quotes.

The principle of causality dictates that $\Delta t = t_2 - t_1$ has the same sign as $\Delta t' = t'_2 - t'_1$, otherwise the order of events would be reversed, the cause would precede the effect; the same reasoning if we imagine the emission and reception of a signal from the point of view of S, the change of sign of the corresponding time interval in S' would indicate that reception precedes emission.

From the above, it follows in particular that:

*if a class T of coordinate transformations provides for the possibility that each pair of events can have opposite temporal orders depending on the inertial frame in which it is represented, then T is incompatible with the principle of causality.*

### §1.5 Some clarifications

Only boosts along the same axis will be considered $X$ and with origin coinciding with a common initial time; in this scheme the only parameter that characterizes an inertial system S' with respect to an inertial system S is the algebraic velocity $v$ of the origin of S' in the system S.

The algebraic velocity v is positive if the motion has the same direction as the X axis and is negative otherwise.

A notion of contemporaneity must be provided even in different spatial positions, this notion, central to the measurements, is operational and common to each SRI, there must be a common standard for time intervals and lengths.

Let us now make two slightly more formal requests, essentially following Cattaneo's analysis [8]:

- for every S and S' every function of an event E that is differentiable in S, where the event is individuated by the real quatern (x,y,z,t), must also be differentiable in S' where the same event is individuated by the real quatern (x',y',z',t');

- the boost from S to S' is differentiable at least twice with continuity on S, invertible and with inverse differentiable at least twice on S' (in other words, the boosts are of class $C^{(2)}$ on S and S' ).

These conditions allow us to construct boosts that are at least locally linear and to imagine, we are reasoning in a purely kinematic field, velocity and acceleration of the "moving points" as continuous functions (for an in-depth study of the themes of differential geometry with applications to physics [27,28]).

## 2. The deduction of the form of possible transformations

## §2.1 Transformations are linear

The event E has coordinates $(x,y,z,t)$ in S and $(x',y',z',t')$ in S', for reasons of formal convenience we put $x_1 \equiv x$, $x_2 \equiv y$, $x_3 \equiv z$, $x_4 \equiv t$ . In general, we have that:

$$x_1' = f_1( x_1, x_2, x_3, x_4, v) \tag{2.1}$$
$$x_2' = f_2( x_1, x_2, x_3, x_4, v)$$
$$x_3' = f_3( x_1, x_2, x_3, x_4, v)$$
$$x_4' = f_4( x_1, x_2, x_3, x_4, v)$$

where $f_k$ are functions of at least class $C^{(2)}$ on an open $A \subseteq R^4$, this condition guarantees a locality character to the proposed developments; the velocity $v$ is at least of class $C^{(0)}$ on an open $\Lambda \subseteq R$.

Let us now impose the postulates of the homogeneity of space and time and consider a particular reference event $E_0$; the same variation of $x_k$ must always have the same effect on the corresponding variation of $x'_k$ regardless of the choice of $E_0$ (for the homogeneity of space and time every point in space-time is equivalent, the properties of bodies are invariant). The isotropy of space then guarantees that directions different from that of the $\vec{v}$ originally chosen vector are equivalent; in particular, if we choose one of the coordinate axes, we will denote it with X, in the new direction of $\vec{v}$ we will obtain the same transformation equations for the other two coordinate axes.

$$x_1' + \Delta x_1' = f_1( x_1 + \Delta x_1, x_2, x_3, x_4) \tag{2.2}$$
$$x_2' + \Delta x_2' = f_2( x_1, x_2 + \Delta x_2, x_3, x_4)$$
$$x_3' + \Delta x_3' = f_3( x_1, x_2, x_3 + \Delta x_3, x_4)$$
$$x_4' + \Delta x_4' = f_4( x_1, x_2, x_3, x_4 + \Delta x_4)$$

With $\Delta x'_k$ depending only on $\Delta x_k$ and the parameter $v$; given the formal analogy we can reason for a generic k, for example k=1, *and then extend the results obtained to the other values of k.*
Combining the first of (2.1) with the first of (2.2) we obtain that:

$$\Delta x_1' = f_1( x_1 + \Delta x_1, x_2, x_3, x_4, v) - f_1( x_1, x_2, x_3, x_4, v)$$

from which dividing both sides by $\Delta x1$ we have:

$$\frac{\Delta x_1'}{\Delta x_1} = \frac{f_1( x_1 + \Delta x_1, x_2, x_3, x_4, v) - f_1( x_1, x_2, x_3, x_4, v)}{\Delta x_1} \tag{2.3}$$

*At this point we can introduce students to a first definition of partial derivation and to the idea of total differential, there is no need to deepen the discourse, it is enough to have an intuitive idea of the mathematical tool. What is assumed is only a good knowledge of the ordinary differential calculus faced in the mathematics courses of the high schools.*
For what has been said, the second member of (2.3) depends only on $\Delta x_1$ and on $v$, which plays the role of a parameter, the same happens for the first member; this continues to be true even if in (2.3) we pass to the limit $\Delta x_1 \rightarrow 0$ .
Moreover, from the assumptions made in §1.5 we know that $f_k$ are at least of class $C^{(2)}$ and then such a limit exists and (2.3) becomes the definition of partial derivative of $f_1$ with respect to $x_1$.

Partial derivative and not total because the variation concerns only one variable, $x_1$, while the others are kept constant; the total derivation operator $d/dx_1$ will be replaced by the partial derivation operator $\partial/\partial x_1$.

We can then say that $\partial f1/\partial x1$ is a constant, the only variable on which it can depend is the parameter $v$, by analogy with the ordinary differential calculus we write that:

$$x'_1 = \frac{\partial f_1}{\partial x_1} x_1 + A_1\big(x_2, x_3, x_4, v\big)$$

where $A_1(x_2, x_3, x_4, v)$ is an expression that contains only $x_2, x_3, x_4, v$ and therefore does not change with the variation of $x_1$. Considering below the partial variations of $x_2, x_3, x_4$ we can write that:

$$x'_1 = \sum_{k=1}^{4} \frac{\partial f_1}{\partial x_k} x_k + C_1(v) \tag{2.4}$$

The overall variation of $x_1$ is its total differential, which we denote by $dx_1$; this variation is given by the sum of the independent contributions due to the partial variations relative to $x_1, x_2, x_3, x_4$ (not of $v$, which is simply a parameter).

$$dx'_1 = \sum_{k=1}^{4} \frac{\partial f_1}{\partial x_k} dx_k \tag{2.5}$$

Extending the discussion to all variables, let's write the transformation equations in a synthetic way:

$$x'_J = \sum_{k=1}^{4} \frac{\partial f_J}{\partial x_k} x_k + C_J(v) \tag{2.6}$$

$$dx'_J = \sum_{k=1}^{4} \frac{\partial f_J}{\partial x_k} dx_k \tag{2.7}$$

Since we are limiting ourselves to boost $TB_0$ for simplicity, the event identified in S by (0,0,0,0) must correspond in S' to the quatern (0,0,0,0); but then in (2.6) we can put $C_J(v)=0$ and move on to the relation:

$$x'_J = \sum_{k=1}^{4} \frac{\partial f_J}{\partial x_k} x_k \tag{2.8}$$

The transformations of coordinates and differentials are linear; In matrix form we have that:

$$X' = A(v) \cdot X \quad e \quad dX' = A(v) \cdot dX \tag{2.9}$$

where X' and X are the column vectors of the stressed and unstressed variables while A(v) is the matrix such that $a_{jk}(v) = \partial f_j/\partial x_k$.

If the transformations were not linear then some $\partial f_j/\partial x_k$ would depend on x and we would have $dx'_j$ function of x.

As an example, let's consider only one spatial dimension and the transformation function:

$x' = ax_2 + bt$ , $t' = cx + kt_2$ ; then $dx' = 2axdx + bdt$ and $dt' = cdx + 2ktdt$ .

The interval between neighbouring points would be transformed differently by changing the reference position in space and time, against the hypotheses of homogeneity of space and time.

The linearity of the transformations has as a consequence that a uniform rectilinear motion in S is transformed into a still uniform rectilinear motion according to S' (see App.3).

*The derivation proposed here has the advantage of assuming only the hypotheses of homogeneity of space and time; There is no reference to the principle of relativity.*

One could arrive in a less general way at the conclusion that transformations are linear also starting from the observation that for the principle of relativity a uniform rectilinear motion in S must be uniform rectilinear for every other inertial frame S' [see App.3] .

## § 2.2 A first set of restrictions on the form of transformations

It may be useful to graphically represent the motion of S' as seen by S.

**Fig.1**
Initial condition for t=t'=0

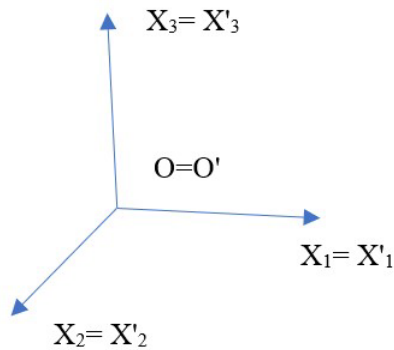

The plane $\alpha(x_2x_3)$ has in S the equation $x_1=0$ and the plane $\alpha(x'_2x'_3)$ has in S' the equation $x'_1=0$. When t=0 the two planes coincide, therefore we deduce that $(x_1=0$ and $t=0) \rightarrow x'_1=0$.

By entering this information in (2.7) we obtain that $0=a_{12}x_2+a_{13}x_3$ . This relation must be true for every possible value of $x_1$ and $x_2$, so $a_{12}$ and $a_{13}$ must both be zero: $a_{12} = a_{13} = 0$

**Fig.2**
situation at time t of S

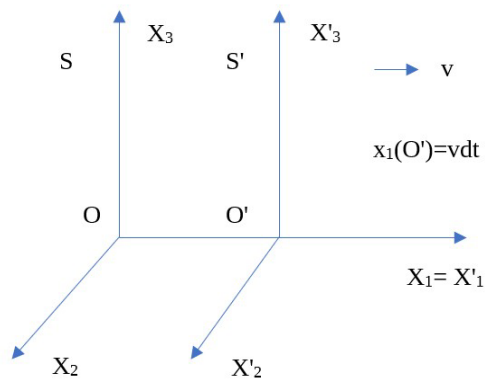

The plane $\alpha(x_1 x_2)$ has in S the equation $x_3=0$ and the plane $\alpha(x'_1 x'_2)$ has in S' the equation $x'_3=0$.
From the point of view of S, the plane $\alpha(x'_1 x'_2)$ represents, for every possible value of $t$, a sliding of $\alpha(x_1 x_2)$ on itself; for every $t$ we have that $(x_3=0) \rightarrow (x'_3=0)$.
In other words, $0= a_{31}x_1 + a_{32}x_2 + a_{34}x_4$ whatever the value of $x_1$, $x_2$, $x_4$; then $a_{31} = a_{32} = a_{34} = 0$.
For the plane $\alpha(x_1 x_3)$ we reason in a similar way: for every t we have that $(x_2=0) \rightarrow (x'_2=0)$, which translates into the relation $0= a_{21}x_1 + a_{23}x_3 + a_{24}x_4$ from which follows $a_{21} = a_{23} = a_{24} = 0$.
If we insert the information obtained into the equations of (2.7) expressing $x'_2$ and $x'_3$ we conclude that $x'_2 = a_{22}x_2$ and $x'_3 = a_{33}x_3$.
The postulate of the isotropy of space implies that the coordinate axes $X_2$ and $X_3$ must be considered equivalent; the only privileged direction is that of $\vec{v}$, velocity of O' with respect to S, which coincides with that of the axis $X_1$. It must then be $a_{22} = a_{33}$, let us say $F(v) = a_{22} = a_{33}$.
Again for the isotropy of space, the transformation of the time coordinate, $x_4$, cannot depend on $x_2$ or $x_3$, in other words $a_{42} = a_{43} = 0$.
At this point, for the sake of clarity, let's explain that $x_4$ is the time variable and formalize the results achieved on the transformation equations:

$$dx'_1 = a_{11}(v)dx_{1 \# xA0;} + a_{14}(v)dt \qquad (2.10)$$

$$dx'_2 = F(v)dx_2$$

$$dx'_3 = F(v)dx_3$$

$$dt' = a_{41}(v)dx_1 + a_{44}(v)dt$$

## §2.3 Still Further Restrictions

Remember that the parameters $a_{jk}(v)$ that appear in (2.10) depend only on v and not from the particular point that is being considered, it follows that if we can understand how a particular point is transformed then we can also draw some information on the shape of the parameters $a_{jk}(v)$. Let us consider in this regard what happens with reference to the origin O' of S'.
Its origin is seen with spatial coordinates (0,0,0) for each time t', then it follows that:
$dx'_1 = dx'_2 = dx'_3 = 0$. S sees O' moving with a uniform rectilinear motion of velocity v along $X_1$, so $dx_1 = vdt$, $dx_2=0$, $dx_3=0$.
The second and third of (2.10) turn into identities and provide no information. The fourth tells us that if $dx_1 = vdt$ then $dt' = a_{44}(v)vdt + a_{44}(v)dt = [a_{44}(v)v + a_{44}(v)]dt$, in this situation dt' and dt are directly proportional.
The first of (2.10) is richer in information, which becomes $0 = a_{11}(v)vdt + a_{14}(v)dt$; since this relation must hold for every dt, we immediately obtain that $a_{14}(v) = - a_{11}(v)v$.
Let's examine what happens if we invert the axes $X_1$ and $X'_1$, let u be the speed that characterizes the transformation. For the hypothesis of isotropy of space the only variation is that $x_1$ becomes $-x_1$ and $x'_1$ becomes $-x'_1$, $dx_1$ becomes $-dx_1$ and $dx'_1$ becomes $-dx'_1$.
S sees the displacement of O' as $dx_1 = vdt$ and when we invert the axes S sees the displacement of O' as $-dx_1 = udt$; it must then be $u = -v$.
After replacing $a_{14}(v)$ with $- a_{11}(v)v$, we move on to the equations of the transformation, taking into account that $-dx'_1 = -a_{11}(-v)dx_1 + a_{11}(-v)vdt$ is equivalent to $dx'_1 = a_{11}(-v)dx_1 - a_{11}(-v)vdt$.

$$dx'_1 = a_{11}(-v)dx_1 - a_{11}(-v)vdt \qquad (2.11)$$

$$dx'_2 = F(-v)dx_2$$

$$dx'_3 = F(-v)dx_3$$

$$dt' = -a_{41}(-v)dx_1 + a_{44}(-v)dt$$

(2.11) must be equivalent to (2.10), then the following conditions apply to the parameters:
$F(v)=F(-v)$ ; $a_{11}(v) = a_{11}(-v)$ ; $a_{44}(v) = a_{44}(-v)$; $a_{41}(v)=-a_{41}(-v)$ .
Remembering that $a_{14}(v) = -a_{11}(v)v$ and therefore $dx'_1 = a_{11}(v)(dx_1 - vdt)$, we can obtain further constraints by imposing that v=0 corresponds to the identity transformation.
From the first three equations of (2.9) we immediately obtain that $a_{11}(0)=F(0)=1$ , from the fourth equation it follows that $dt' = a_{41}(0)dx_1 + a_{44}(0)dt = dt$ for every possible $dx_1$ and therefore $a_{41}(0)=0$ and $a_{44}(0)=1$ .

## §2.3 Invertibility condition, parameter sign

Let's rewrite (2.11) inserting the information obtained so far:

$$dx'_1 = a_{11}(v)dx_1 - a_{11}(v)vdt \qquad\qquad (2.11_{Bis})$$

$$dx'_2 = F(v)dx_2$$

$$dx'_3 = F(v)dx_3$$

$$dt' = a_{41}(v)dx_1 + a_{44}(v)dt$$

With $a_{11}(0)=F(0)= a_{44}(0) = 1$, $a_{41}(0)=0$ ; $F(v)$, $a_{11}(v)$, $a_{44}(v)$ even functions of v ; $a_{41}(v)$ odd function of v.
By hypothesis the functions $a_{jk}(v)$ are continuous, in particular $a_{11}(v)$ is also continuous and for it the Bolzano theorem is applicable, from which it follows that if $a_{11}(v)$ changed sign then there should exist a value v* for which $a_{11}(v^*) = 0$.
But $a_{11}(v)\neq 0$ and therefore the sign of $a_{11}(v)$ is constant; moreover, we know that $a_{11}(0)=1$, we deduce that $a_{11}(v) > 0$ .
Transformations must be invertible, so the determinant of the matrix associated with them is non-zero. We then observe that the first and fourth of $(2.11_{Bis})$ are independent of the second and third; In essence, it is permissible to move on to two systems of equations whose only link is constituted by the parameter v.

$$dx'_1 = a_{11}(v)dx_1 - a_{11}(v)vdt \qquad (2.12) \qquad\qquad dx'_2 = F(v)dx_2 \qquad (2.13)$$

$$dt' = a_{41}(v)dx_1 + a_{44}(v)dt \qquad\qquad\qquad\qquad dx'_3 = F(v)dx_3$$

The invertibility condition of (2.12) is equivalent to the requirement that the determinant D of the corresponding matrix be non-zero: $D=a_{11}(v)\cdot a_{44}(v) + v\cdot a_{11}(v)\cdot a41(v) \neq 0$ and therefore $a_{11}(v)\neq 0$ and also $a_{44}(v) + v\cdot a_{41}(v) \neq 0$ . The invertibility condition of (2.13) is $F(v) \neq 0$ .
Let's consider the second of (2.12), if $dx_1=0$ then $dt' = a_{44}(v)dt$ and invertibility is guaranteed only if $a_{44}(v)\neq 0$. From $a_{11}(0)=1$, repeating the same analyses seen for $a_{11}(v)$, we deduce that $a_{44}(v) > 0$ .
Inverting (2.13) we get $dx_2 = [1/F(v)]dx'_2$ and $dx_3 = [1/F(v)]dx'_3$ . For the principle of relativity we must have form identity with (2.13) and therefore also $dx_2 = F(u)dx'_2$ and $dx_3 = F(u)dx'_3$ where u is the parameter, to be determined, that characterizes the transformation from S' to S, that is, the speed with which S' sees the origin O of S moving. In essence, it must be $F(u)=[1/F(v)]$ and therefore $F(u)F(v)=1$.

As far as u is concerned, we can draw some indications directly from (2.12) by observing that the speed u with which S' sees the origin O of S moving can be obtained by substituting $dx_1=0$ for each relation and dividing the first by the second of (2.12).

$$dx'_1 = - a_{11}(v)vdt \qquad (2.14)$$

$$dt' = a_{44}(v)dt$$

$$u = \frac{dx'}{dt'} = - \frac{a_{11}(v)}{a_{44}(v)}$$

Let's now reverse (2.12):

$$dx_1 = \frac{a_{44}(v)}{a_{11}(v)\left[a_{44}(v) + a_{41}(v)v\right]}dx'_1 + \frac{a_{11}(v)v}{a_{11}(v)\left[a_{44}(v) + a_{41}(v)v\right]} \qquad (2.15)$$

$$dt = \frac{- a_{41}(v)}{a_{11}(v)\left[a_{44}(v) + a_{41}(v)v\right]}dx_1' + \frac{a_{11}(v)}{a_{11}(v)\left[a_{44}(v) + a_{41}(v)v\right]}dt'$$

By the principle of relativity, (2.15) must be identical in form to (2.12) when they are rewritten in terms of the parameter u, which leads to the formulation of the following four conditions

$$a_{11}(u) = \frac{a_{44}(v)}{a_{11}(v)\left[a_{44}(v) + a_{41}(v)v\right]} \qquad (A1)$$

$$-ua_{11}(u) = \frac{a_{11}(v)v}{a_{44}(v) + a_{41}(v)v} \qquad (A2)$$

$$a_{41}(u) = \frac{- a_{41}(v)}{a_{11}(v)\left[a_{44}(v) + a_{41}(v)v\right]} \qquad (A3)$$

$$a_{44}(u) = \frac{1}{a_{44}(v) + a_{41}(v)v} \qquad (A4)$$

## §2.4 A demonstration of the reciprocity lemma

In addition to the hypotheses of homogeneity and isotropy of space, homogeneity of time and the principle of relativity, the relations A1 , A2 , A3 , A4 have been deduced from simple considerations of linear algebra and allow us to reach the conclusion that u=-v, the so-called *Reciprocity lemma*, in an alternative way with respect to some characteristic derivations of the literature on the subject [6,7,8] .

Dividing member by member A2 by A1 we still get the relation (2.14):

$$u = -\frac{a_{11}(v)}{a_{44}(v)} \cdot v$$

Dividing A3 by A4 gives that:

$$\frac{a_{41}(u)}{a_{44}(u)} = \frac{-a_{41}(v)}{a_{11}(v)}$$

The equivalence in form of A1 , A2, A3 , A4, for the exchange of u with v , guaranteed by the principle of relativity, allows us to deduce the relation :

$$\frac{a_{41}(v)}{a_{44}(v)} = \frac{-a_{41}(u)}{a_{11}(u)} \qquad (2.15_B)$$

From $(2.15_A)$ and $(2.15_B)$ we derive two expressions for $a_{41}(v)$, after some algebraic manipulation we arrive at the equation:

$$\frac{a_{11}(v)}{a_{44}(v)} = \frac{a_{44}(u)}{a_{11}(u)} \qquad (2.15_C)$$

There must then exist a constant K for which :

$$\frac{a_{11}(v)}{a_{44}(v)} = K \quad \text{and} \quad \frac{a_{44}(u)}{a_{11}(u)} = K \; ; \text{the last relation is equivalent to} \quad \frac{a_{11}(u)}{a_{44}(u)} = \frac{1}{K} \; .$$

By the principle of relativity, the function $a_{jk}(u)$ is identical in form to the function $a_{jk}(v)$; we can then state that $(1/K) = K$, or $K^2 = 1$. There are therefore only two possibilities: K=1 or K=1. But in §2.3 we had demonstrated that $a_{11}$ and $a_{44}$ are positive definite, then we conclude that K=1. In other terms we can then say that $a_{44}(v) = a_{11}(v)$.
Substituting this relation in the second of (2.14) we obtain that u=-v .
We have thus demonstrated in a simple way the so-called **Lemma of reciprocity** :

*If v is the speed with which the inertial system S sees the inertial system S' move then S' sees S moving at velocity -v .*

## §2.5 Towards the general form of transformations

The condition $u=-v$ has significant consequences on the form of F(v). We have already obtained that F(u)F(v)=1 and therefore we can write F(-v)F(v)=1, from which, since F(v) is even, $F^2(v)=1$ follows. We deduce that for any value of $v$ or F(v)=-1 or F(v)=1. After all, we know that F(0)=1 and then it must be F(v)=1 . At this point, substituting in (2.13), we definitively formulate the transformation relations for $x_2$ and $x_3$ .

$$dx'_2 = dx_2 \qquad (2.16)$$

$$dx'_3 = dx_3$$

Now let's use A4 to arrive at a form that contains only one function $a_{jk}(v)$.

Substituting $a_{11}(v)$ for $a_{44}(v)$ in A4 we obtain $a_{11}(v)(a_{11}(v)+a_{41}(v))=1$ from which, in the hypothesis that $v \neq 0$, we express $a_{41}(v)$ in terms of $a_{11}(v)$:

$$a_{41}(v) = -\frac{1}{v} \cdot \frac{a_{11}^2(v) - 1}{a_{11}(v)} \tag{2.17}$$

Apparently, the discontinuity for $v=0$ seems to exclude the identity transformation, which is physically obtained precisely for $v=0$.

The difficulty is easily overcome if the following condition applies:

$$\lim_{v \to 0} a_{41}(v) = 0 \tag{2.18}$$

In this case, by definition putting $a*_{41}(v)=a_{41}(v)$ if $v \neq 0$ *and* $a*_{41}(v)=0$ if $v=0$, we obtain the identity transformation at $v=0$; in order not to weigh down the notation, we will write $a_{41}(v)$ instead of $a*_{41}(v)$. It is also clear that the necessary verification of the validity of (2.18) can only be done a posteriori once $a_{11}(v)$ has been determined, in this regard reference is made to paragraph § 3.1 and Appendix 4 . In view of a formulation of equations similar to those widespread in the literature for Lorentz transformations, we adopt the following conventions:

$$\gamma(v) \equiv a_{11}(v) \text{ and } \varphi(v) \equiv -\frac{a_{41}(v)}{a_{11}(v)} = \frac{a_{11}^2(v) - 1}{v a_{11}(v)}$$

Given the analysis made for $a_{41}(v)$ we should also introduce a function $\varphi*(v)$ analogous to $a*_{41}(v)$, in order not to weigh down the notation we will still write $\varphi(v)$ instead of $\varphi*(v)$.

Starting from (2.12), taking into account (2.17) and the conventions just adopted, we arrive at the general form of the transformation equations:

$$dx'_1 = \gamma(v)(dx_1 - vdt) \tag{2.19}$$
$$dt' = \gamma(v)(dt - \varphi(v)dx_1)$$

To be considered together with (2.16).

# 3. Ignatowsky's Constant

## 3.1 Three inertial systems

Let S, S' and S" be three inertial systems such that when $t=t'=t"=0$ we have $O=O'=O"$ and $X_k=X'_k=X"_k$. S sees S' move according to a speed boost $v_1$ along $X_1$, let $A(v_1)$ be the relative transformation; S" sees S" moving according to a speed boost $v_2$ along $X_1$, let $A(v_2)$ be the relative transformation.

We want to determine the velocity $v_3$ at which S sees S" moving; the idea is to go from S to S" via S' following the scheme $S \to S' \to S"$ with S' being the transform of S .

S' → S": boost A($v_2$).

$$dx''_1 = \gamma(v_2)\left(dx'_1 - v_2 dt'\right)) \qquad (3.1)$$
$$dt'' = \gamma(v_2)\left(dt' - \varphi(v_2)dx'_1\right)$$

where $dx'_1$ and $dt'$ are the transforms of dx and dt with the boost A($v_1$):

$$dx'_1 = \gamma(v_1)\left(dx_1 - v_1 dt\right) \qquad (3.2)$$
$$dt' = \gamma(v_1)\left(dt - \varphi(v_1)dx_1\right)$$

Substituting in (3.1) we obtain after a few algebraic steps that:

$$dx''_1 = \gamma(v_1)\gamma(v_2)\left[\left(1 + v_2\varphi(v_1)dx_1 - \left(v_1 + v_2\right)dt\right)\right] \qquad (3.3)$$
$$dt'' = \gamma(v_1)\gamma(v_2)\left[\left(1 + v_1\varphi(v_2)\right)dt - \left(\varphi(v_1) + \varphi(v_2)\right)dx_1\right]$$

We observe that O is characterized in S by dx=0 for every dt and therefore its corresponding in S", i.e. how S" sees O, is obtained by placing dx=0 in (3.3).

$$dx''_1 = \gamma(v_1)\gamma(v_2)\left[-\left(v_1 + v_2\right)dt\right] \qquad (3.3\text{-bis})$$
$$dt'' = \gamma(v_1)\gamma(v_2)\left[1 + v_1\varphi(v_2)dt\right]$$

Dividing the first by the second of (3.3-bis) we have the speed with which S" sees S move:

$$\frac{dx''_1}{dt''} = -\frac{v_1 + v_2}{1 + v_1\varphi(v_2)}$$

and then changing the sign, the one, $v_3$, with which S sees S" move:

$$v_3 = \frac{v_1 + v_2}{1 + v_1\varphi(v_2)} \qquad (3.4)$$

Furthermore, according to the principle of relativity, the form of (3.3) must coincide with that of (3.1) and therefore $\gamma(v_3) = \gamma(v_1)\gamma(v_2)[1+v_2\,\varphi(v_1)]$ and also $\gamma(v_3) = \gamma(v_1)\gamma(v_2)[1+v_1\,\varphi(v_2)]$; equalizing the two second members of these relations we obtain that for each velocity pair ($v_1$, $v_2$) must be:

$$v_2\,\varphi(v_1) = v_1\,\varphi(v_2) \qquad (3.5)$$

If one of the two velocities is zero, then instead of $\varphi$ we consider the corresponding $\varphi^*$, which is also zero according to §2.5; for example, if $v_2 = 0$ then $\varphi^*(v_2)=0$, then both sides of (3.5) become equal to zero and the relation is verified. It is therefore not restrictive to suppose $v_1$, $v_2 \neq 0$.

Whatever $v_1$, $v_2 \neq 0$ are, we can reason in general terms and consider that for a generic velocity v there are only two possibilities:
(a) [($\varphi(v) = 0$) for each $v \neq 0$]
b) ($\varphi(v) \neq 0$) for each $v \neq 0$].
If $\varphi(v) = 0$ then, as we have defined $\varphi$, must be $\gamma^2(v) = 1$; but $\gamma(v) > 0$ and therefore $\gamma(v) = 1$.
Inserting $\varphi(v) \neq 0$ e $\gamma(v)=1$ in (2.19) and remembering that $dx'_2 = dx_2$ and $dx'_3 = dx_3$ we get:

$$dx'_1 = dx_1 - vdt$$

$$dx'_2 = dx_2 \tag{3.6}$$

$$dx'_3 = dx_3$$

$$dt' = dt$$

The (3.6) coincide with Galileo's transformations.
In particular, we observe that $(\varphi(v) = 0) \rightarrow dt'=dt$ which means invariance of time intervals.
Let us now impose in (2.19) the constraint dt'=dt for every value of v and $dx_1$ ; consider the second of (2.19): dt' = $\gamma(v)(dt - \varphi(v)dx_1)$ . The condition dt'=dt implies $\gamma(v)=1$ and $\varphi(v) = 0$ .
We have thus shown that (dt'=dt)$\rightarrow$($\varphi(v) = 0$), but we know that ($\varphi(v)= 0$)$\rightarrow$(dt'=dt) and then the relation (dt'=dt)$\Leftrightarrow$($\varphi(v) = 0$) is also proved.
Now let's go back to the fundamental (3.5), from the non-restrictive hypothesis $v_1$, $v_2 \neq 0$ we get that

$$\frac{\varphi(v_2)}{v_2} = \frac{\varphi(v_1)}{v_1} \tag{3.7}$$

Since the ratios that appear in (3.7) do not depend on the velocity parameter, then the $\varphi$ratio $(v)/v$ must be a universal constant.
In other words:
$\varphi(v)$ must have the form $\varphi(v) = \alpha v$ where $\alpha$ is a universal constant; even the case examined above, $\varphi(v)=0$, can be included in the form just given by simply putting $\varphi(v) = \alpha v$ with $\alpha = 0$.
Let us now move on to the final formulation of the coordinate transformation equations. We have already examined the case in which $\alpha=0$ obtaining the Galilean transformations, then consider only the cases $\alpha<0$ and $\alpha >0$.
Equations (2.19) become:

$$dx'_1 = \gamma(v)(dx_1 - vdt)$$

$$dt' = \gamma(v)(dt - \varphi(v)dx_1) \tag{2.19_A}$$

Taking into account the conventions adopted previously we have $\gamma(v) \equiv a_{11}(v)$ and also:

$$\varphi(v) = \frac{a^2_{11}(v) - 1}{va^2_{11}(v)} = \frac{\gamma^2(v) - 1}{v\gamma^2(v)}$$

Substituting $\alpha v$ for $\varphi(v)$ we get the equation in $\gamma(v)$: $\gamma^2(v)\alpha v_2 = \gamma^2(v)$ -1 . Therefore it must be:

$$\gamma^2(v) = \frac{1}{1 - \alpha v^2}$$

But $\gamma^2(v)$ is positive and therefore $(1-\alpha v_2)>0$ .

At this point it is easy to demonstrate, see Appendix 4, that

$$\lim_{v \to 0} a_{41}(v) = 0$$

and therefore the discontinuity of the transformations for v=0 can be eliminated; we can properly speak of identity transformation at v=0.

As far as the transformation of velocities is concerned, substituting in (3.4) $\alpha v$ instead of $\varphi(v)$ we obtain:

$$v_3 = \frac{v_1 + v_2}{1 + \alpha v_1 v_2} \tag{3.8}$$

In which $v_2$ is the velocity of a point P with respect to S', $v_1$ is the velocity of S' with respect to S and $v_3$ is the velocity of P with respect to S.

If $\alpha=0$ we naturally obtain the Galilean transformation of the velocities: $v_3 = v_1 + v_2$.

Case $\alpha<0$ .

Condition $(1-\alpha v_2)>0$ is automatically satisfied for every possible value of $v$.

Assuming $\alpha=-1/k^2$ and remembering the condition $a_{11}(v) \equiv \gamma(v)>0$ we can write that:

$$\gamma(v) = \gamma(v) = \frac{1}{\sqrt{1 + \frac{v^2}{k^2}}}$$

All values of v are allowed, there is no speed limit.

The coordinate transformation equations become:

$$dx'_1 = \frac{dx_1 - vdt}{\sqrt{1 + \frac{v^2}{k^2}}}$$

$$dx'_2 = dx_2 \tag{3.9}$$

$$dx'_3 = dx_3$$

$$dt'_1 = \frac{dt + \frac{v}{k^2} dx_1}{\sqrt{1 + \frac{v^2}{k^2}}}$$

and the speed transformation becomes:

$$v_3 = \frac{v_1 + v_2}{1 + \frac{v_1 v_2}{k^2}} \qquad (3.10)$$

Case $\alpha > 0$.

The condition $(1 - \alpha v^2) > 0$ is not automatically satisfied for every possible value of $v$; in fact it must be $v^2 < 1/\alpha$ and therefore $-\frac{1}{\sqrt{\alpha}} < v < \frac{1}{\sqrt{\alpha}}$.

$V_L = \frac{1}{\sqrt{\alpha}}$ then plays the role of limiting speed, we can therefore write that $|v| < V_L$.

Remembering that $a_{11}(v) \equiv \gamma(v) > 0$ we get:

$$\gamma(v) = \frac{1}{\sqrt{1 - \frac{v^2}{V_L^2}}}$$

The coordinate transformation equations become:

$$dx'_1 = \frac{dx_1 - v\,dt}{\sqrt{1 - \frac{v^2}{V_L^2}}} \qquad (3.11)$$

$$dx'_2 = dx_2$$

$$dx'_3 = dx_3$$

$$dt'_1 = \frac{dt - \frac{v}{V_L^2}\,dx_1}{\sqrt{1 - \frac{v^2}{V_L^2}}}$$

The velocity transformation equation becomes:

$$v_3 = \frac{v_1 + v_2}{1 + \frac{1}{V_L^2} v_1 v_2}$$

In summary, there are only three possibilities:

**(i)** $\alpha = 0$

$$v_3 = v_2 + v_1$$

The velocity modulus has no upper limit

**(ii)** $\alpha < 0$

$$v_3 = \frac{v_1 + v_2}{1 - \frac{1}{k^2} v_2 v_1}$$

The velocity modulus has no upper limit

**(iii)** $\alpha > 0$

$$v_3 = \frac{v_1 + v_2}{1 + \frac{1}{V_L^2} v_2 v_1}$$

Speed is limited,
$|v| < V_L$.
$V_L$ plays the role of speed limit

## § 3.2 Possible values of Ignatowsky's constant and form of transformations

W. A. von Ignatowsky [2] was the first to show in 1910 the existence of the universal constant $\alpha$, which we will then call Ignatowsky's constant; we analyse below the consequences of the three possible cases for the Ignatowsky constant.

**(i)** $\alpha = 0$

In the previous paragraph we have shown that (dt'=dt) $\Leftrightarrow (\varphi(v) = 0)$; in terms of $\alpha$ :

(dt'=dt) $\Leftrightarrow (\alpha = 0)$.

A rich series of experimental studies excludes the invariance of the measurement of time intervals in the transition between two inertial frames. Very cogent experimental results have been obtained in the last fifty years starting from the works of J.C.Hafele and R.E.Keating [ref.29] and C.Alley [30], involving atomic clocks in flight, and those relating to the decay of unstable particles, see the review work of J.Bayley [31] ; for a brief exposition of the state of the art see for example G. Giuliani [32].

Excluding $\alpha = 0$ also means excluding the relative law of transformation of velocities, which does not provide for any limiting velocity.

**(ii)** $\alpha < 0$

There are three arguments against this possibility, the first being a kind of kinematic paradox:

if S sees S' moving at positive velocity v1 and S' sees S" moving at positive velocity v2 then S can see S" moving at negative velocity v3 (for details see App.3); this is a conclusion that, although paradoxical, is nevertheless admissible.

The second argument is stronger, the set of coordinate transformations does not constitute a group. In particular, there are infinite pairs (v2,v1) for which there is a singularity in the transformation equations; in these cases S sees S" moving at infinite speed (for details see App.3 ).

At this point $\alpha < 0$ could already be excluded, there is however the violation of the principle of causality as a third and fundamental argument.

We fix our attention on the equation of transformation of time intervals and multiply the first and second members by

$$\frac{1}{\sqrt{1+(v/k)^2}}$$

which is positive definite .

Indicating the sign function with sgn($\cdot$) we have that sgn(dt')=sgn(dt+($v$/k2)dx)=sgn[dt(1+uv/k2)], where $u=dx$/dt and dx is in S the spatial distance of two events separated by a time interval dt.

Imagine, for example, that in the system S the displacement of a particle in a time interval dt is dx≠0, if in the system S' the series of events is seen with sgn(dt') = - sgn(dt) then the temporal order of the events is reversed with respect to S.

The condition for which the principle of causality holds is the existence of at least one set of values of u for which sgn(dt') = sgn(dt) over the whole set of SRI; this is true if and only if (1+uv/k$^2$)>0 for every possible value of $v$.

In other words, if any value of u is fixed, there is always some value of $v$ for which (1+uv/k$^2$)<0 then the principle of causality is violated; and this is precisely what happens.

Solving the inequality (1+uv/k$^2$)<0 in v we have uv<-k$^2$ and there are two possible cases: u>0 or u<0. If u>0 then the solution is v<-k$^2$/u and therefore there are infinite SRI, i.e. infinite values of v, in which the order of events is reversed.

If u<0 then the solution is v<-k$^2$/u and again there are infinite SRI, i.e. infinite values of v, in which the order of events is reversed. According to what is stated in §1.4 we have a violation of the principle of causality.

**(iii)** $\alpha$>0

The set of speeds is limited: $\left| v \right| < V_L$ , with $V_L$>0. With respect to the operation of composition, the set of transformations constitutes a group, see for example [33,35,36] ; Here we focus on the verification of the closure property using simple algebraic methods. To this end, let's start from (3.12) which provides the speed $v_3$ of the boost by composing two boosts, one of speed $v_1$ and the other of speed $v_2$ with $\left| v_1 \right|$ , $\left| v_2 \right| < V_L$ . The closing property holds if and only if $\left| v_3 \right| < V_L$ .

$$v_3 = V_L^2 \, \frac{v_1 + v_2}{V_L^2 + v_1 v_2}$$

The condition $\left| v_3 \right| < V_L$ is the conjunction of the two relations a) $v_3 < V_L$ and b) $v_3 > - V_L$ .

The verification of a) is equivalent to solving the inequality:

$$V_L^2 \cdot \frac{\left( v_1 + v_2 \right)}{V_L^2 + v_1 v_2} < V_L \quad \text{, that is, the inequality} \quad \frac{V_L \left( v_1 + v_2 \right)}{V_L^2 + v_1 v_2} < 1$$

From $\left| v_1 \right|$ , $\left| v_2 \right| < V_L$ follows $V_L^2 + v_1 v_2$ >0, then $V_L (v_1 + v_2) < V_L^2 + v_1 v_2$ from which follows $v_2(V_L - v_1) < V_L(V_L - v_1) \rightarrow v_2 < V_L$ which is always true because $\left| v_2 \right| < V_L$ .

In case b) we proceed in a similar way. The verification of a) is equivalent to solving the inequality:

$$V_L^2 \cdot \frac{\left( v_1 + v_2 \right)}{V_L^2 + v_1 v_2} > - V_L \quad \text{, that is, the inequality} \quad \frac{V_L \left( v_1 + v_2 \right)}{V_L^2 + v_1 v_2} > -1$$

$V_L (v_1 + v_2) > (-V_L^2 - v_1 v_2) \rightarrow V_L (v_1 + V_L) > -v_2(V_L + v_1)$ .

Since $v_1 + V_L > 0$ we conclude that $V_L > -v_2 \leftrightarrow -V_L < v_2$ which is always true because $|v_2| < V_L$.

No kinematic paradox arises, in fact if $v_1$ and $v_2$ have the same sign then from $V^2_L + v_1 v_2 > 0$ and from the fact that

$$v_3 = V^2_L \frac{v_1 + v_2}{V^2_L + v_1 v_2}$$

it is deduced that $\operatorname{sgn}(v_3) = \operatorname{sgn}(v_1 + v_2)$.

Finally, we verify consistency with the principle of causality. Proceeding in a similar way to what we have already seen for $\alpha < 0$ we have that $\operatorname{sgn}(dt') = \operatorname{sgn}(dt)$ if and only if $(1 - uv/VL2) > 0$.

By choosing u in such a way that $|u| < V_L$ and remembering that $|v| < VL$, we obtain that the relation $(1 - uv/V^2_L) > 0$ is verified for every permissible value of v. The principle of causality applies.

From the analyses carried out it is clear that, in the framework of the assumed postulates, the choice $\alpha > 0$ is the only admissible one; the form of the transformations of coordinates and velocities is analogous to those of special relativity, it is sufficient to substitute $V_L$ instead of c.

### § 3.3 Special relativity as a special case

Put $\alpha > 0$ is equivalent to assuming the existence of a limiting speed, which allows us to reformulate the results achieved starting from the following postulates:

1. *The laws of physics have the same form in all inertial frames of reference (principle of relativity).*
The frame of reference consisting of the homogeneity of space-time, the isotropy of space and more generally what is presented in section I of this work is assumed.

2. *There is a limiting velocity $V_L$ that has the same value in all inertial reference frames.*
Postulate 2 also represents a theoretical synthesis of the experimental results that exclude the invariance of the measurement of time intervals in the transition between two inertial frames (see §3.2), an invariance that is instead predicted by Galilean transformations for which there is no limiting velocity.

Special relativity is obtained by adding the following third postulate:
3. *$V_L = c$, speed of light in a vacuum.*

The general theory deduced on the basis of the principles of symmetry expressed by postulates 1 and 2 represents a framework within which to place the physical laws and cannot by its nature give indications on the actual value of $V_L$.
In the final analysis, it is the very broad experimental confirmation of Maxwell's theory of electromagnetism that supports the introduction of the third postulate and with it special relativity.

## Conclusions

A fundamental point of discontinuity in the transition from classical physics to the general theory introduced here is the abandonment of the idea of instantaneous action at a distance in favour of that of the existence of a maximum speed $V_L$ for the propagation of interactions; the variation in the measurement of time intervals in the transition between two inertial frames is an indirect confirmation of the existence of $V_L$.
The break with the Newtonian scheme is qualitatively contained in the proposed scheme, in order to obtain quantitative evaluations it is necessary to specify the measure of $V_L$. Special relativity provides for this by assuming the validity of Maxwell's equations, i.e. the invariance of the speed of light in a vacuum.

In the course of the discussion we obtained as an intermediate result an algebraic proof of the reciprocity lemma that is simpler than other approaches proposed in the literature [6,7,8].

We conclude by quoting some passages from the analysis that L.D. Landau makes in the first paragraph of his classic "Field Theory" [34].

*"Experience shows, however, that there are no instantaneous interactions in nature... In reality, if one of the interacting bodies undergoes some change, the repercussion on another body in the system will occur after a certain interval of time".*

*"From the principle of relativity it follows, in particular, that the speed of propagation of interactions is the same in all inertial frames of reference. The speed of propagation of interactions is therefore a universal constant."*

## Thanks

I thank Giuseppe Giuliani, Biagio Buonaura and Achille Cristallini for their valuable suggestions and critical attention; I thank Claudia Maria Belardi who with her observations has contributed to improving the drafting of this work.

## Appendix 1

It is easy to prove that *if* there is a limiting speed $V_L$ and c ,speed of light in a vacuum, is constant *then* $V_L = c$ .

Following any manual of relativity [16 ] it is easy to show how from the existence of a limiting velocity $V_L$ follows the well-known rule of transformation of velocities:

$$u = \frac{v + u'}{1 + \frac{vu'}{V_L^{\ 2}}}$$

where v is the velocity of the inertial frame S' with respect to the inertial frame S, u' is the velocity of a particle with respect to the inertial frame S', u is the velocity of the particle with respect to S. We are reasoning in the hypothesis that v and u' have the same direction.

Let's assume that there is a constant velocity c; for each u' must then be:

$$c = \frac{c + u'}{1 + \frac{cu'}{V_L^{\ 2}}} = V_L^2 \frac{u' + c}{V_L^{\ 2} + u'c}$$

from which we have $c(V_L^2 + cu') = cV_L^2 + u' V_L^2$ , then $c^2 u' - u' V_L^2 = 0$ and then $c^2 = V_L^2$ which implies $\left| V_L \right| = \left| c \right|$ .

## Appendix 2

If Ignatowsky's constant $\alpha$ is negative then any value of the velocity is allowed, but problems arise from the application of the formula for the transformation of velocities.

Remembering what has been obtained in paragraph 3.2, we can write that:

$$v_3 = \frac{v_1 + v_2}{1 - \frac{v_1 v_2}{k^2}} = k^2 \frac{v_1 + v_2}{k^2 - v_1 v_2} \tag{A2.1}$$

where $v_1$ is the velocity of the inertial system S' with respect to the inertial system S, $v_2$ is the velocity of the inertial system S" with respect to the inertial system S', $v_3$ is the velocity of S" with respect to S. We are thinking in the hypothesis that $v_1$ and $v_2$ have the same direction.

In the plane $(v_2, v_1)$ we then have a singularity whenever we find ourselves on the equilateral hyperbola $v_1 v_2 = k^2$.

In this case v3 diverges and S sees S" moving at infinite speed, this condition has no physical sense; we are not able to directly connect S" to S by reasoning within the news.

The set of TLG transformations is not closed for the internal composition operation $\cdot$ :

there are infinite pairs $(v_2, v_1)$ for which $T_{LG}(v_2) \cdot T_{LG}(v_1) \notin T_{LG}$ ; it follows in particular that $T_{LG}$ cannot be a group.

This is an important argument against the possibility of it $\alpha$ being negative.

In addition to this, it is necessary to underline a question which, although admissible, appears paradoxical: if S sees S' moving at positive velocity $v_1$ and S' sees S" moving at positive velocity $v_2$, then S can see S" moving at negative velocity $v_3$.

From (A21) it follows that if v>0 and u>0 then the sign of w coincides with that of $k^2 - uv$.

But k2 – uv <0 ↔uv> $k^2$ and therefore we conclude that:

$( v>0 , u>0$ and $uv> k^2 ) \rightarrow v_3 < 0$

# Appendix 3

From the principle of relativity it follows that a uniform rectilinear motion in S must be uniform rectilinear for every other inertial frame S', a necessary condition is that the transformations of the coordinates transform lines of S into lines of S' or are [16 , pp. 19-20] "conformal" of the type:

$$x'_1 = \frac{ax_1 + bx_2 + cx_3 + dx_4 + E}{\alpha x_1 + \beta x_2 + \gamma x_3 + \delta x_4 + F} \tag{A3.1}$$

Similar formulas naturally apply to x'2 , x'3 , x'4 .

But transforming lines into lines is not enough to pass from a uniform rectilinear motion to another uniform rectilinear motion, finite points must be mapped to finite points and infinite points to infinity points.

Taking (A.31) as a reference, if at least one of the parameters $\alpha, \beta, \gamma, \delta$ is non-zero, the points of the line r) of the equation $\alpha x_1 + \beta x_2 + \gamma x_3 + \delta x_4 + F = 0$ represent a singularity for the transformation. They are transformed into points at infinity unless they are points in common with the line m) $ax_1 + bx_2 + cx_3 + dx_4 + E = 0$ .

In this case r) and m) have only one point in common and therefore there are still infinite singularities or r)=m), In the latter case we can eliminate the singularity by putting by definition x'1 = G, with G constant,

This transformation makes us pass from an open A of $R_4$ to an open of $R_3$, the hyperplane $x'_1 = G$, but in doing so we lose the possibility that the transformation is one-to-one from A to A.

From what has been said we conclude that the only admissible situation is that parameters $\alpha, \beta, \gamma, \delta$ are all zero and then the denominator of (A.31) is reduced to F, which obviously must be different from zero. (A.31) is equivalent to :

$$x'_1 = a_{11}x_1 + a_{12}x_2 + a_{13}x_3 + a_{14}x_4 + C_1$$

Generalizing:

$$x'_j = \sum_{k=1}^{4} a_{jk}x_k + C_j(v) \tag{A3.2}$$

which coincides with (2.6) of §2.1; transformations are linear.

## Appendix 4

In §3.1 we have shown that $\gamma^2(v) = \dfrac{1}{1 - \alpha v^2}$ and from $a_{11}(v) = \gamma(v)$ follows:

$$a_{41}(v) = -\frac{1}{v}\frac{a_{11}(v) - 1}{a_{11}(v)} = -\frac{1}{v}\frac{\alpha v^2}{\dfrac{1}{\sqrt{1 - \alpha v^2}}} = \alpha v\sqrt{1 - \alpha v^2}$$

From this it immediately follows:

$$\lim_{v \to 0} a_{41}(v) = 0$$